\documentclass{IEEEtran}
\usepackage{cite}
\usepackage{amsmath,amssymb,amsfonts}
\usepackage{graphicx}
\usepackage{textcomp,nicefrac}
\usepackage{subfig}
\usepackage{float}
\usepackage{color}
 \usepackage{booktabs}
 \usepackage{multirow}
 \usepackage{graphicx}
 \usepackage{enumitem}

\def\BibTeX{{\rm B\kern-.05em{\sc i\kern-.025em b}\kern-.08em
T\kern-.1667em\lower.7ex\hbox{E}\kern-.125emX}}
\begin{document}
    
\title{Experimental Findings on the Sources of\\ Detected Unrecoverable Errors in GPUs} 

\author{Fernando Fernandes dos Santos+, Sujit Malde*,
    Carlo Cazzaniga*\\Christopher Frost*, Luigi Carro+, and Paolo Rech+\\
    +Institute of Informatics of Universidade Federal do Rio Grande do Sul (UFRGS), Brazil\\
    *Science and Technology Facility Council (STFC), UKRI
\thanks{
This project has received funding from the European Union's Horizon 2020 research and innovation
programme under the Marie Sklodowska-Curie grant agreement No 886202 and from The Coordena\c{c} \~{a}o de Aperfei\c{c}oamento de Pessoal de N\'{i}vel Superior - Brazil (CAPES) - Finance Code 001. 
}
}



    \maketitle
    \begin{abstract}
We investigate the sources of Detected Unrecoverable Errors (DUEs) in GPUs exposed to neutron beams. Illegal memory accesses and interface errors are among the more likely sources of DUEs. ECC increases the launch failure events.
Our test procedure has shown that ECC can reduce the DUEs caused by Illegal Address access up to 92\% for Kepler and 98\% for Volta.

    \end{abstract}


%

\section{Introduction}
\label{sec_introduction}

Graphics Processing Units (GPUs) have evolved from graphics rendering to general-purpose accelerators extensively employed in HPC and safety-critical applications such as autonomous vehicles for the automotive and aerospace markets. 
The highly parallel architecture of GPUs fits the computational characteristic of most HPC codes and of Convolutional Neural Networks (CNNs) used to detect objects. The most recent GPU architecture advances, such as tensor core and mixed-precision functional units, move toward improving the performances and software flexibility for HPC and deep learning applications.

Today, the reliability of parallel processors is a significant concern for both safety-critical applications and HPC systems. Unexpected errors in parallel devices' may lead to catastrophic accidents in self-driving vehicles and, in HPC systems, to lower scientific productivity, lower operational efficiency, and even significant monetary loss.

Most recent studies target Silent Data Corruption (SDC) in their evaluation. SDCs, being undetectable, are in fact considered the main threat for modern computing devices reliability~\cite{Baumann2005}. Detected Unrecoverable Errors (DUEs), such as device hangs, application crashes, or functional interruptions, are considered less harmful as, being detectable by definition, they could be easily handled using solutions such as checkpoints, and software/hardware watchdogs~\cite{cao2016CheckpointAndRestart,chen2007SoftwareWatchdog}. 
Nevertheless, the recovery from a DUE or the action taken to reach a fail-safe state require a significant amount of time, which risks reducing supercomputers productivity. 
A small cluster with 32K cores would take almost an hour to restart after a crash~\cite{cao2016CheckpointAndRestart}, without considering the overhead of performing checkpointing time.
In safety-critical real-time systems, such as autonomous vehicles, the DUE risk is even higher, as it may compromise the system's ability to complete the task before the deadline. For instance, a GPU for autonomous vehicles must process 40 frames-per-second. The recovery from a DUE must be sufficiently efficient not to miss any frame, which is highly challenging. In this scenario, tracing the software and hardware sources for DUEs and quickly identify the occurrence of a DUE are an essential tools to create more tolerant applications against crashes and hangs.

In this paper, we investigate the sources of DUE in two NVIDIA architectures: Kepler and Volta. 
We provide a novel and detailed analysis of DUE sources on GPUs, based on neutron experimental data and system logs profile. We create a framework that allows the tracing of the GPU crashes and hangs observed during radiation experiments.
We select a set of eight algorithms and compare their DUE and SDC rates, considering both the case of ECC disabled and enabled.
Each code has peculiar characteristics regarding memory utilization, computing power, control-flow operation, highlighting
specific architecture behaviors that could be generalized to similar algorithms. We report findings from recently completed (remotely controlled) neutron beam testing that represents a total of more than 2 million years of operation in a natural environment. Finally, we discuss how the use of system log tracing can make DUEs detection (and thus recovery) faster.

\section{Radiation induced SDCs and DUEs in GPUs} 
\label{section_background}

A transient fault leads to one of the following outcomes: (1) no effect on the program output (i.e., the fault is masked, or the corrupted data is not used), (2) a Silent Data Corruption (SDC) (i.e., an 
incorrect program output), or (3) a Detected Unrecoverable Error (DUE) (i.e., a program crash or device reboot).
Previous studies have stated that parallel architectures, particularly GPUs, have a high fault rate because of the high amount of available resources~\cite{Nathan2014,
DATE2014}. Recent works have identified some peculiar reliability weaknesses of GPUs architecture, suspecting that the corruption of the GPU hardware scheduler or shared memories can severely impact the computation of several parallel threads~\cite{Nathan2014, 
 Oliveira2017, tc2016}. 
As a result, multiple GPU output elements can potentially be corrupted, effectively undermining several applications' reliability, including CNNs~\cite{Santos_TR, Ibrahim2020}.

Even if DUEs are detectable, they can lead to monetary loss or harmful events. For instance, a self-driving car that relies on a GPU to perform object detection, if rebooted, can delay a response to a critical situation, thus putting human lives in danger. DUEs are generated by data-driven control-flow errors (corruption of a memory address, an index, a jump instruction, etc.) or by faults in control logic (scheduler, memory controller, interfaces for synchronization, etc.). Additionally, the available Single Error Correction Double Error Detection (SECDED) ECC crashes the application when a a double bit flip is detected.
Our paper's goals are: (1) understand the (software/hardware) causes of DUE in GPU. DUEs events are expected to be generated in other places than functional units~\cite{Lunardi2018}. As we show, DUEs come from incorrect memory addresses, errors in the scheduler, errors in the stack, etc. (more details Section~\ref{section_due_tracing}). (2) Show how GPU system logs can be used to have prompt information about DUE occurrence, triggering  the action to take faster. 
\section{GPU DUE source tracing} 
\label{section_due_tracing}

While to detect SDC is sufficient to compare the experimental output with the pre-computed expected one, to identify the event that triggered the DUE it is necessary to query the device or the operating system to understand what happened. GPUs feature the CUDA Runtime Application Programming Interface (CUDA Runtime API), a layer between the operating system and device hardware. With CUDA Runtime API, we can directly control the GPU using function calls. For instance, we can make a software reset on the GPU (i.e., delete all memory and kernels resources) by calling \textit{cudaDeviceReset()}. 

We used the CUDA Runtime API to trace and log the DUE events in our experiments. 
In the following, we list and describe the possible DUEs we have identified. 

	
\noindent \textbf{Devices Unavailable:} the GPU is unavailable either as it is in Exclusive or Prohibited mode or the previous execution has not released the resources (the GPU is "falsely" full).
\\
\noindent \textbf{Illegal Instruction:} an illegal instruction is executed, leaving the process in an inconsistent status.  
\\
\noindent \textbf{Illegal or Misaligned Address:} the address of a Load or Store does not point to a valid memory address or is not aligned.                 
\\
\noindent \textbf{Initialization Error:} the CUDA driver fails to initialize, making it impossible for the API to continue.  
\\
\noindent \textbf{Invalid Device or No Device:} the GPU queried by the CPU is invalid, or the CUDA driver detects no device.                                      \\               
\noindent \textbf{Invalid PC:} one of the kernel program counters is 
invalid.                                     \\	
\noindent \textbf{Invalid Address Space:} a kernel can operate in different memories spaces (i.e., global, shared, or local). This error happens when an instruction that belongs to a specific memory space tries to operate in a different one.
\\
\noindent \textbf{Memory Allocation:} CUDA is not able to allocate memory.
\\
\noindent \textbf{ECC Uncorrectable:} the ECC detects more than one bit flip. As this cannot be corrected, the ECC throws an exception.
\\
\noindent \textbf{Launch Failure:} the CPU tries to launch a kernel and fails. There are many reasons that it could happen, such as dereferencing an invalid device pointer and accessing out of bounds shared memory, etc.
\\	
\noindent \textbf{Hardware Stack Error:} an error in the call stack during kernel execution, generally due to stack corruption or exceeding the stack size limit.
\\	
\noindent \textbf{Invalid Value:} some kernel parameters are out of the acceptable range of values, i.g., more threads than the GPU supports.
\\
\noindent \textbf{System Crash:} a DUE triggered by the watchdog in our setup. Then DUE source could not be traced or is generally unknown.

\begin{table}[]
	\caption{Benchmarks used for reliability evaluation}
	\label{tab_benchmarks}
	\resizebox{.49\textwidth}{!}{%
		\begin{tabular}{lll}
			\hline
			\textbf{}                                        & \textbf{Domain}                       & \textbf{Suite}                              \\ \hline
			\multicolumn{1}{l|}{Component Labeling - CCL}    & \multicolumn{1}{l|}{Graph}            & NUPAR~\cite{nupar2015}                      \\
			\multicolumn{1}{l|}{Breadth First Search - BFS}  & \multicolumn{1}{l|}{Graph}            & \multirow{6}{*}{Rodinia~\cite{rodinia2009}} \\
			\multicolumn{1}{l|}{Lava}                        & \multicolumn{1}{l|}{N-Body}           &                                             \\
			\multicolumn{1}{l|}{Gaussian}                    & \multicolumn{1}{l|}{Linear Algebra}   &                                             \\
			\multicolumn{1}{l|}{LU Decomposition - LUD}      & \multicolumn{1}{l|}{Linear Algebra}   &                                             \\
			\multicolumn{1}{l|}{Matrix Multiplication - MXM} & \multicolumn{1}{l|}{Linear Algebra}   &                                             \\
			\multicolumn{1}{l|}{Optimized MXM - GEMM}        & \multicolumn{1}{l|}{Linear Algebra}   &                                             \\
			\multicolumn{1}{l|}{Mergesort}                   & \multicolumn{1}{l|}{Sorting}          & NVIDIA SAMPLES                              \\
			\multicolumn{1}{l|}{Quicksort}                   & \multicolumn{1}{l|}{Sorting}          & NVIDIA SAMPLES                              \\
			\multicolumn{1}{l|}{YOLOv3}                      & \multicolumn{1}{l|}{Object Detection} & Darknet~\cite{yolov3}                       \\ \hline
		\end{tabular}%
	}
\end{table}

\section{Evaluation Methodology}
\label{section_methodology}



\textbf{Devices:} We consider Kepler (Tesla K40) and Volta (Titan V and Tesla V100) NVIDIA GPUs. 
The tested K40~(\textbf{Kepler}) is fabricated in a $28nm$ TSMC standard CMOS technology. 
SECDED ECC protects the register file, shared memory and caches.
The Titan V and Tesla V100 (\textbf{Volta}) are built with TSMC FinFET 12nm. 
Volta GPUs support three IEEE754 float point precisions, double, float, and half, plus
eight \textit{tensor cores}, i.e., specific hardware that performs $16x16$ Matrix Multiplication. 
For Kepler, we chose a beam spot sufficiently small (2cm of diameter) not to hit the onboard DDR when ECC is disabled.  For Volta, as HBM2 memories are on top of the chip when ECC is disabled, all the global memory accesses are made through Triple Modular Redundancy (TMR). 


\textbf{Tested Codes:}
We chose the eight representative codes listed in Table~\ref{tab_benchmarks}, from HPC and deep learning domains. The choice of diverse codes increases this work's quality and can provide details for DUEs extendable to different applications~\cite{errorBarHeather2014}.
Because of its importance in CNNs and HPC, we choose to pay particular attention to Matrix Multiplication and test both the naive version (\textit{MxM}) and the optimized version that digest data in the most suitable way for GPUs as General Matrix Multiplication (\textit{GEMM}) from the NVIDIA CUBLAS libraries. To be highly efficient, GEMM kernel is tuned for each input and device configuration. 
Float based codes have their precision in their names. D for double, F for float, and H for half. 
INT32 based codes do not have their names modified.


%
%


\begin{figure}[t]
	\centering{
		\includegraphics[width=0.98\linewidth]{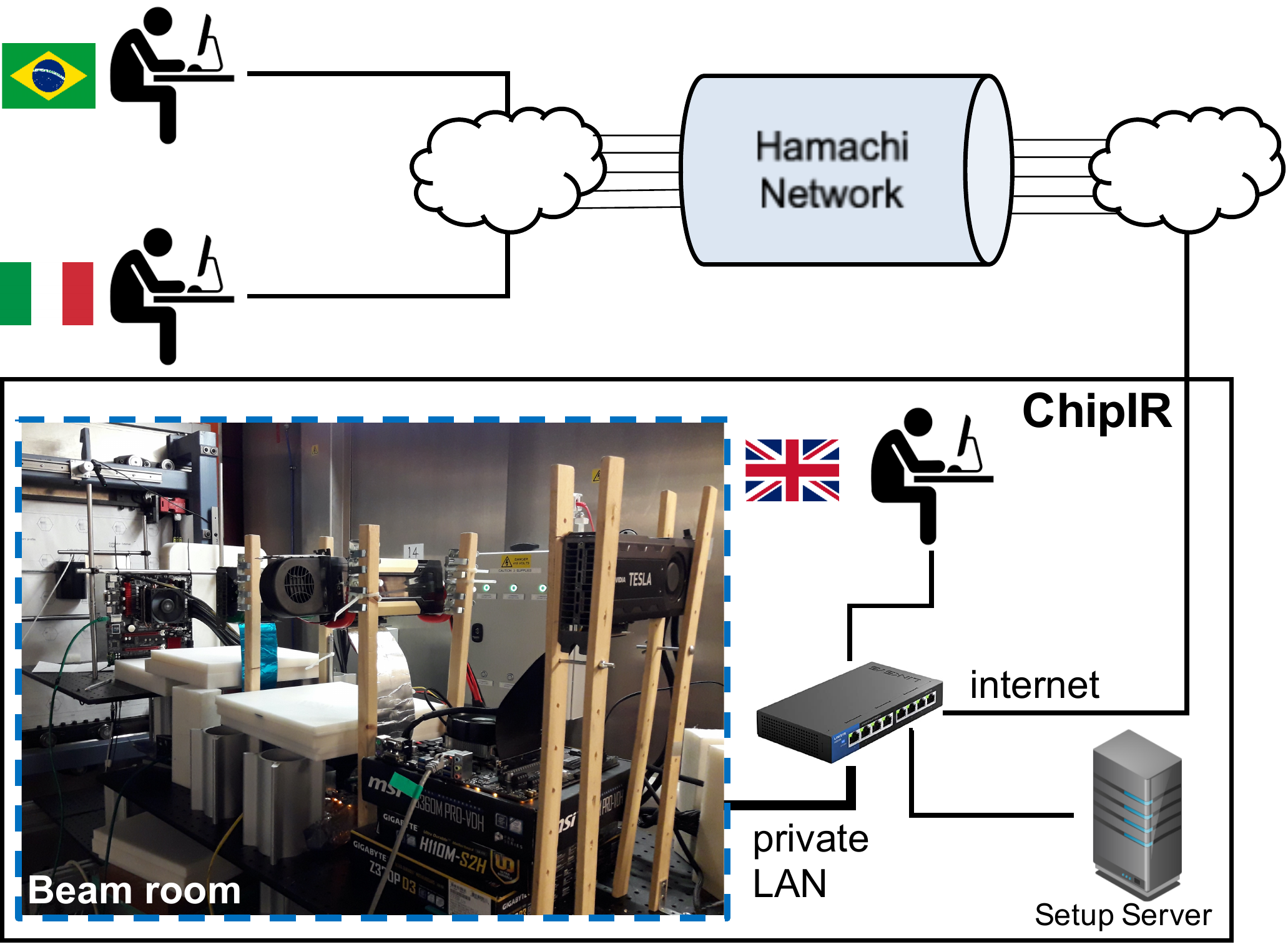}
		\caption{Remotely controlled beam experiment setup}
		\label{fig_setup_covid}
	}
\end{figure}
   
\textbf{Beam Experiment Setup:} our experiments were performed at the ChipIR facility of the Rutherford Appleton Laboratory, UK. Figure~\ref{fig_setup_covid} shows the setup mounted in the ChipIR facility. The facility delivers a beam of neutrons with a spectrum of energies that resembles the atmospheric neutron one~\cite{Cazzaniga_2018}.
The available neutron flux was about $3.5\times10^6 n/(cm^{2}/s)$. 

Most experiments were performed during the \textit{COVID-19 pandemic}. This demanded the setup to be adapted for remote control, since not all the scientists involved in this research were physically present at the experiments. Figure~\ref{fig_setup_covid} shows the adapted setup for mobility restrictions imposed by the COVID-19. A Virtual Private Network was created using Hamachi LogMeIn to connect researchers in Brazil, Italy, and United Kingdom (ChipIR). A researcher was in charge of assembling the hardware setup on the ChipIR facility and configuring the network. The other researchers were then able to remotely monitor and adjust the irradiated GPUs' applications. 

\begin{figure*}[]
	\centering
	\includegraphics[width=\linewidth]{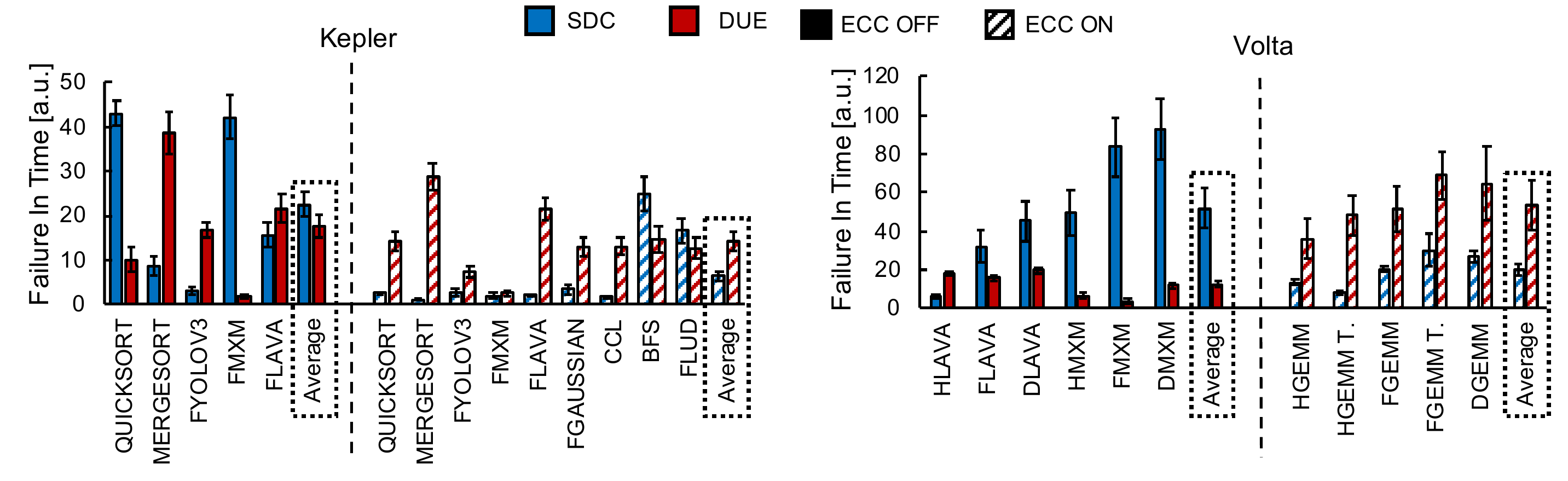}
	\caption{Normalized FIT rates for Kepler and Volta.}
	\label{fig_beam}
\end{figure*}

We added a software and a hardware watchdog to recover from crashes. The software watchdog monitors the application under test, and if it stops responding in a predefined time interval, the kernel is killed and relaunched. This watchdog detects kernel crashes or software hangs, i.e., application crashes or control flow errors that prevent the GPU from completing assigned tasks (e.g., an infinite loop).
Before finishing the process related to the GPU's kernel, our setup makes a CUDA API call to get the last error that happens inside the GPU. Then, the return of the API call is logged in a structured file.
The hardware watchdog is an Ethernet-controlled switch that performs a host computer's power cycle if the host computer itself does not acknowledge any ping requests in a predefined time interval. The hardware watchdog is necessary to detect when the operating system hangs.


\section{Neutron Beam Experiments Results}
\label{section_fit_beam}

In this section we give an overview of the SDC and DUE experimental data. In the next section we detail the identified DUE sources. In the final paper, fault injection will be added to track faults propagation and better identify errors sources.

Figure~\ref{fig_beam} shows the experimentally measured SDC and DUE \textit{normalized FIT} rates for the GPUs executing the codes with ECC disabled and enabled. We report arbitrary units (a.u.) not to reveal business-sensitive data.
Values are reported with 95\% confidence intervals considering a Poisson distribution.

Not surprisingly, both the SDC and DUE FIT rates of Kepler and Volta change significantly depending on the code.
It is interesting to notice that, on the \textit{Average}, the SDC FIT rate is higher than the DUE rate when ECC is OFF (1.2$\times$ for Kepler, 4.1$\times$ for Volta),
while when the ECC is ON the DUE rate is (significantly) higher than the SDC rate (2.2$\times$ for Kepler, 2.7$\times$ for Volta). 
This is because the ECC reduces (significantly) SDCs and increases DUEs, mostly because of ECC Uncorrectable errors (details in Section~\ref{section_due_tracing}). For Kepler, the average SDC FIT rate with ECC OFF is up to 21$\times$ higher than with ECC ON. The DUE FIT rate, when ECC is ON, increases of up to 1.4$\times$ on Kepler and 13.7$\times$ on Volta.  
The DUE increase caused by ECC is exacerbated in Volta GPUs, because of the main memory stacked on top of the chip and thus being irradiated as the GPU core. 
It is worth noting that, when the ECC is OFF, we triplicate data in the DDR to avoid its corruption to bias our SDC evaluation.


For Mergesort and YOLOv3 (object detection CNN), on Kepler, even with ECC OFF, the DUE rate is 4.4$\times$ and 5.1$\times$ higher than the SDC rate. Mergesort continuously requires the GPU to exchange data with the CPU, as the array split is performed in the host. Similarly, YOLOv3, which is a neural network of more than 100 layers, launches several kernels per layer and requires a high amount of data to go through the bus (each layer output is used as input in the downstream layer), which increases the probability of invalid device pointers to be corrupted leading to a DUE. The synchronizations between CPU and GPU and heavy memory exchange are also highly susceptible, and their corruption is likely to cause a DUE. 
As a counter-proof, Quicksort, despite solving the same sorting problem (with the same input) as Mergesort, uses \textit{Dinamic Paralelism}, which allows launching kernel inside the GPU without requiring data exchange with the CPU. As a result, Quicksort has a much higher SDC rate than DUE rate. The fact that Quicksort always uses the GPU increases its SDC rate when ECC is OFF (that is 4.9$\times$ higher than Mergesort SDC rate) but, when ECC is ON, most SDCs are corrected, making Quicksort more reliable than Mergesort. 

For Volta GPU, we also perform the analysis for FIT rates of codes executed with different precisions: 64bit (D), 32bit (F), and 16bit (H). A higher precision functional unit has a higher area and a higher probability of being hit by a neutron since the circuit area is bigger. 
Thus, higher precision implies a higher number of bits to store data, which has a linear dependence with the FIT rate. The SDC FIT rates for Volta in Figure~\ref{fig_beam} confirm this trend.

When ECC is OFF, the DUE rate is almost constant for the different precisions (while it changes between Lava and MxM). This is because the probability of faults in the control circuit, interfaces, host-device communications (that are likely to generate a DUE) does not depend on the data precision. 
Interestingly, when ECC is ON, the DUE FIT rate increases linearly with data precision. This is again due to the DDR being stacked over the GPU chip. In fact, the higher the amount of data, and consequentially, the higher the probability of having ECC uncorrectable errors. This is confirmed by the analysis of the DUE sources we present in the next Section. 

\section{DUE source}
\label{section_due_source}

\begin{figure*}[]
	\centering
	\includegraphics[width=\linewidth]{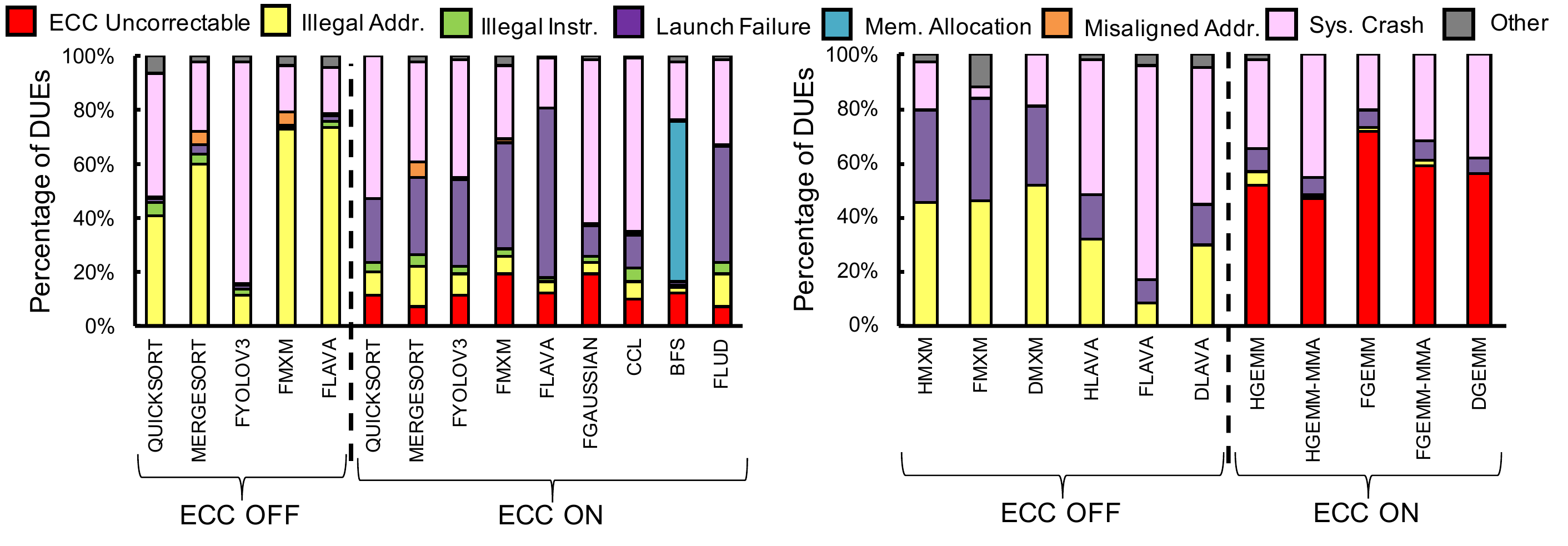}
	\caption{Detailed DUE sources for Kepler and Volta GPUs. Other Sources includes Devices Unavailable, Invalid Value, No Device, Initialization Error, Hardware Stack Error, and Invalid Device.}
	\label{fig_due_fit}
\end{figure*}

Thanks to the setup described in Section~\ref{section_due_tracing}, we are able to trace the cause of the observed DUEs.
Figure~\ref{fig_due_fit} shows, in percentage, the sources of DUEs we have identified in our beam experiments, for Kepler and Volta.
In Figure~\ref{fig_due_fit} we have grouped Device Unavailable, Invalid Value, No Device, Initialization Error, Hardware Stack Error, and Invalid Device events under the category "Other" as, on average, the combination of these events caused less than 3\% of DUEs in our experiments.

Figure~\ref{fig_due_fit} shows that DUEs' sources are code and architecture dependent. 
When ECC is disabled, most DUEs are originated by \textit{Illegal Address} (i.e., the code tries to access an incorrect memory address), which is to be expected since the Register File (RF), shared, and cache memories are unprotected, including the ones that store the memory addresses for Load/Store instructions.
A corruption in the memory address is likely to violate the memory policy leading to an Illegal Address. When ECC protects memories, there is a drastic reduction of Illegal Address DUEs (less than 8\% on Kepler and less than 2\% on Volta) and an expected increase in the probability of \textit{ECC Uncorrectable} errors (which are absent when ECC is off). The DUE rate is exacerbated on Volta when ECC is enabled as DDR is on top of the GPU. 
The combination of transient, permanent, and intermittent errors in the stacked DDR makes double-bit errors probability very high (details on DDR errors will be provided in the final paper).




\textit{Launch Failures} are also frequent, mainly for Kepler with ECC on and for Volta.
These DUEs happen when the GPU is in an inconsistent state due to corrupted parameters or an error at the kernel launch. For instance, an error that invalidates the memory pointers passed to the kernel as parameters, making the GPU unable to launch it. 

A \textit{System Crash} source is hard to identify as this DUE is generated by exceptions from the host operating system or hanging kernels inside the device, preventing logging API data. In a System Crash event, our watchdogs kill the application, reset the GPU, or perform a system power-cycle. As the System Crash percentage is not negligible, it would be necessary to investigate their causes deeper.
As future work, we plan to consider the operating system logs related to the PCI bus and the GPU for an in-depth study of System Crashes. 

Finally, BFS is the only code experiencing a large number of \textit{Memory Allocation} DUEs. BFS is a code that manages the GPU memory inefficiently. BFS allocates a graph with 1 million vertices for each CUDA parallel stream (i.e., a CUDA stream is an instance of a BFS kernel that runs parallel with other kernels). 50 CUDA streams will be launched, and for each stream, there will be a high memory demand and irregular memory accesses. Thus, it is expected that BFS will have the majority of the DUEs coming from memory errors.

In all the DUEs we have seen but System Crash, for which the only solution is a power cycle or a device reset, the existing tools allow to get data from the device to have prompt information about the DUE occurrence. Tracing logs can be used to design codes that can self-detect and self-recover from DUEs. For instance, NVIDIA Management Library gives details about ECC Uncorrectable errors. 
This information could be used not to crash the application and triggering a new data fetch or a roll-back instead.
We believe that GPU system logs can be used as DUE detection as already done in other techniques that provide certification and code changes to improve the software reliability~\cite{scade2006}.
A software engineering methodology for safety-critical applications can define actions to be performed soon after a traceable DUE occurs but before the watchdog triggers the device or system reboot. This may save precious time.


%

\section{Conclusions}
\label{section_conclusions}
We have discussed the causes of DUEs for various benchmarks to precisely evaluate the GPU behavior under radiation. 
Although ECC is a powerful technique to reduce the SDC rate, at the same time it can drastically change how the GPU manifests the DUE events. In future research, we plan to deeply study the DUEs cause by ECC and find the causes of System Crashes.

\bibliographystyle{IEEEtran}
\bibliography{IEEEabrv,refreduced}

\end{document}